\documentstyle[12pt]{article}
\newcommand{\tline}{\setlength{\baselineskip}{0.84cm}}

\begin{document}
\tline
\tline
%\spapar
\setcounter{page}{1}
\begin {center}
{\large {\bf Reply to Comment on ``On the Original Proof by Reductio}}\\
{\large {\bf ad Absurdum of the Hohenberg-Kohn Theorem}}\\
{\large {\bf for Many-Electron Coulomb Systems"}} \\
{\large {\bf [Szczepanik, W; Dulak, M.; Wesolowski, T. A.}}\\
{\large {\bf Int J Quantum Chem 2006, 106, .]}}
\end{center}
\vspace{0.25cm}
\begin{center} 
Eugene S. KRYACHKO\footnote[1]{E-mail address: eugene.kryachko@ulg.ac.be}
\end{center}
\vspace{0.25cm}
{\normalsize \centerline{Bogoliubov Institute for Theoretical Physics, Kiev, 03143 Ukraine}}
{\normalsize \centerline{and}}
{\normalsize \centerline{Department of Chemistry, Bat. B6c, University of Liege}}
{\normalsize \centerline{Sart-Tilman, B-4000 Liege 1, Belgium}}
\vspace{0.5cm}
{\normalsize \centerline{ {\em Submitted to:}}} 
{\normalsize \centerline{ {\em The International Journal of Quantum Chemistry}}}
\vspace{0.5cm}
{\normalsize \centerline{PACS number(s): 31.15.Ew, 03.65.Ge, 31.15.Pf, 31.15.-p, 71.15.Mb, 71.15.Qe}
\vspace{1cm}

Any mathematical proof is a game. As a game, it is based on a definite set of rules of logic reasoning which altogether constitutes the subject of logic.

One of the simplest rules of the theory of logic is a denial of the truth of a given proposition that is expressed as a sentence. The truth of a proposition has to be denied by asserting its negation. Assuming, for example, that the proposition p :=$\; \{$Everyone is wise$\}$ (see Ref. [1]) is false, I assert instead s :=$\; \{$Everyone is unwise$\}$. s seems to be a negation of p. However, s is definitively not the logical negation of p that in logic is defined as `not p' or ${}^\sim$p := $\{$the proposition that is true when p is false and false when p is true$\}$ [1,2]. Therefore, s is false if p is true, but not certainly true if p is false, and hence, s $\neq {}^\sim$p.    

The standard method of asserting the negation of a simple sentence consists in attaching the word `not' to the main verb of the sentence. However, the assertion of the negation of a compound proposition (sentence) is not that trivial. Consider the set of negations within the context of the Hohenberg-Kohn theorem [3] (see also Ref. [4] and Ref. [5] for the notations):
\begin{eqnarray}
\hspace{6cm} &\mbox{Negation}& \nonumber \\
\mbox{p} := \{v_1({\bf r}) - v_2({\bf r}) = \mbox{constant}\} &\Rightarrow&  {}^\sim \mbox{p} := \{v_1({\bf r}) - v_2({\bf r}) \neq \mbox{constant}\} \nonumber \\
\mbox{q} := \{\rho_o^{(1)}({\bf r}) = \rho_o^{(2)}({\bf r})\} \hspace{1.5cm} &\Rightarrow&  {}^\sim \mbox{q} := \{\rho_o^{(1)}({\bf r}) \neq \rho_o^{(2)}({\bf r})\}. 
\label{1}
\end{eqnarray}

According to Ref. [6], the Hohenberg-Kohn negation is the compound proposition ${}^\sim\mbox{p} \& \mbox{q}$ expressed as the conjunction (indicated by the ampersand sign `\&' [2]) of the two simple propositions, ${}^\sim\mbox{p}$ and $\mbox{q}$. The proposition ${}^\sim\mbox{p} \& \mbox{q}$ is true iff both these simple propositions are true, i. e.,    
\begin{equation}
{}^\sim \mbox{p} := \{v_1({\bf r}) - v_2({\bf r}) \neq \mbox{constant}\} \; \mbox{and}\; \mbox{q} := \{\rho_o^{(1)}({\bf r}) = \rho_o^{(2)}({\bf r})\}\; \mbox{are both true}. 
\label{2}
\end{equation}
Conversely, the proposition ${}^\sim\mbox{p} \& \mbox{q}$ is false iff both these simple propositions are false, that is,    
\begin{equation}
{}^\sim \mbox{p} := \{v_1({\bf r}) - v_2({\bf r}) \neq \mbox{constant}\} \; \mbox{and}\; \mbox{q} := \{\rho_o^{(1)}({\bf r}) = \rho_o^{(2)}({\bf r})\}\; \mbox{are both false}. 
\label{3}
\end{equation}
Within the context of Ref. [6], what is proved in the original {\it reductio ad absurdum} (:= {\it RAA}) proof by Hohenberg and Kohn [3] is that ${}^\sim\mbox{p} \& \mbox{q}$ is false. Since the negation of a false proposition is true and since ${}^\sim {}^\sim \mbox{p} = \mbox{p},\; \forall\; \mbox{p}$, the true proposition is actually $\mbox{p} \& {}^\sim\mbox{q}$ which is 
\begin{equation}
\mbox{p} := \{v_1({\bf r}) - v_2({\bf r}) = \mbox{constant}\} \; \mbox{and}\; {}^\sim \mbox{q} := \{\rho_o^{(1)}({\bf r}) \neq \rho_o^{(2)}({\bf r})\}. 
\label{4}
\end{equation}
Absurd: (4) does not have anything in common with the Hohenberg-Kohn {\it RAA} proof. Therefore, the interpretation of the Hohenberg-Kohn theorem presented in Ref. [6] is the authors' own interpretation that is unsatisfactory. 

Actually, Hohenberg and Kohn [3] assert that if the antecedent proposition (hypothesis or premise) ${}^\sim \mbox{p}$ is true then the consequent proposition $\mbox{q}$ is false. In logic, this is precisely the conditional proposition or implication [1,2]. In the other words, the Hohenberg-Kohn implication ${}^\sim \mbox{p} \Rightarrow \mbox{q}$ is false. Within the logic theory [2], it means that this `implication is not truth-functional connective'. 

Any {\it RAA} proof has its own strict rules mainly based on the theory of logical interference [2]. These rules do not allow too much room in exercising of refuting a given proposition that contrasts with the suggestion in Ref. [6]. The application of these rules to the {\it RAA} proof of the Hohenberg-Kohn theorem is demonstrated in Ref. [7]: assume a given premise ${}^\sim \mbox{p}$ for the class of many-electron Coulomb systems and, using the rules of logical derivation, infer those propositions which logically follow from ${}^\sim \mbox{p}$, particularly that, ${}^\sim \mbox{q}$, provided by the Kato electron-nuclear cusp theorem for many-electron Coulomb systems [8]. These propositions as being authentically true if the premise is true form the set ${\mathcal{P}}_1$. ${}^\sim \mbox{p}$ is the negation of the proposition $\mbox{q}$ invoked to build the Hohenberg-Kohn {\it RAA} implication and therefore, these two propositions are contradictory to each other (Ref. [2], p. 37). Hence, ${\mathcal{P}}_1$ is intrinsically incompatible or `inconsistent' (Ref. [2], p. 36ff) with the to-be-refuted premise $\mbox{q}$. It merely implies that the to-be-refuted premise cannot be chosen arbitrarily as being a priori contradictory or incompatible with the set ${\mathcal{P}}_1$. 

It is worth finally mentioning a couple of other, less important shortcomings of Ref. [6]. One is its insufficient logical reasoning. Let us read: ``The author (of Ref. [5]) notes that, as the consequence of the Kato cusp condition [3], the external potential can be uniquely reconstructed from the information contained in the electron density for a Coulomb system. ... However, this observation cannot be used to question the validity of the original reductio ad absurdum (RAA) proof." Why? And read further: ``The logical outline of the two proofs ... makes it evident that the Kato cusp condition can be used to demonstrate that the negation of the Hohenberg-Kohn theorem is false." The another one is related to the fact that the existence of a certain relationship between the one-electron densities and external potentials based on the Kato theorem has been already noticed in the beginning of the eightees by A. J. Coleman, A. S. Bamzai and B. M. Deb, V. H. Smith Jr.${}^\dagger$, and E. Bright Wilson (see Refs. [8-11] in Ref. [5] and also Ref. [4]).  

I gratefully thank all the reviewers and the readers of my early works [5,7] for lively discussions and invaluable suggestions. I also thank the authors of Ref. [6] for sending me reprint of their work prior to its publication and mutual exchanging with them my works [5,7], and would like to express a certain curiosity of their ignorance to apply in Ref. [6] the well established logical rules of the {\it RAA} proof, shortly outlined in my work [7], that would certainly help me to avoid the publication of the present reply. Concluding, I would like to quote E. C. G. Boyle who once stated that ``it makes all the difference in the world whether we put Truth in the first or in the second place."

\end{document}